\documentclass[aps,prl,preprint,groupedaddress]{revtex4-1}

\usepackage{amsfonts}
\usepackage{mathrsfs}
\usepackage{amsthm}
\usepackage{graphicx}

\begin{document}


\title{Detecting the Solution Space of Vertex-Cover by Mutual-determinations and Backbones}




\author{Wei Wei}
\author{Renquan Zhang$^{*}$}
\author{Binghui Guo}
\author{Zhiming Zheng}
\affiliation{LMIB and School of mathematics and systems sciences, Beihang University, 100191, Beijing, China\\ $^{*}$zhangrenquan09@smss.buaa.edu.cn}


\date{\today}

\begin{abstract}To solve the combinatorial optimization
problems especially the minimal Vertex-cover problem with high
efficiency, is a significant task in theoretical computer science
and many other subjects. Aiming at detecting the solution space of
Vertex-cover, a new structure named mutual-determination between
unfrozen nodes is defined and discovered for arbitrary graphs, which
results in the emergence of the strong correlations. Based on the
backbones and mutual-determinations with node ranks by leaf removal,
we propose an Mutual-determination and Backbone Evolution Algorithm
to achieve the reduced solution graph, which provides a graphical
expression of the solution space of Vertex-cover. By this algorithm,
the whole solution space and detailed structures such as backbones
can be obtained strictly when there is no leaf-removal core on the
graph. Compared with the current algorithms, the
Mutual-determination and Backbone Evolution Algorithm performs
better than the replica symmetry ones but has a small gap higher
than the replica symmetric breaking ones, and has a relatively small
error for the exact results. The algorithm with the
mutual-determination provides a new viewpoint to solve Vertex-cover,
by which all detailed information of the ground/steady states can be
shown in the reduced solution graph.


\end{abstract}


\maketitle

\section{Introduction}
The minimal vertex-cover problem (Vertex-cover) belongs to one of
Karp's 21 NP-complete problems \cite{karp} and the six basic
NP-complete problems \cite{cook,NP}, which is considered as one of
the classical problems in theoretical computer science. The aim of
this problem is to mark a minimum subset of vertices such that there
are at least one vertex of each edge in the subset. There are a
large number of applications of this problem in the related real
networks, such as immunization strategies in networks \cite{immu}
and monitoring of internet traffic \cite{moni}.

There is a threshold behavior of the minimum vertex-cover problem on
the Erd$\ddot{o}$s-R$\acute{e}$nyi random graph \cite{random}. It
means that the typical running time of algorithms changes to
exponential from polynomial when the order parameter becomes lager
than the Euler number $e$ \cite{martin1, martin2}. This phase
transition phenomenon is considered to have intrinsic correspondence
with the clustering structure of solution space which have already
been observed in statistical physics when studying spin glasses
\cite{spin,mezard1}. Although most statistical physicists believe
that the clustering structure leads to the failure of replica
symmetry, the details of the relation between searching solutions
and the structure are not well established, and how the clustering
structure looks like is far from being clear for most models
\cite{wolf,maswei,ptwei}. From an algorithmic point of view, the
solutions' structure makes great effects on the algorithm to find
the solutions, which sets barriers to local searching algorithms and
makes the computation expensive \cite{AI,algorithms}. So the
features of solutions' structure are explored by different
approaches. Till now, some typical structures such as clustering,
backbone, backdoor \cite{backbone} and frustration \cite{zhou}, have
been widely investigated to understand the structure of solutions
more clearly. Especially, H. Zhou \cite{zhou,frus} has proposed the
\emph{long-range frustration} structure and F. Krzakala \cite{PNAS}
has provided a formal definition as \emph{long-range correlation}.
The long-range correlation and bakbone structures are treated as the
origin of the replica symmetric breaking and the high computational
complexity. And, based on the analysis of these typical
characteristics of the solution space, many efficient searching
algorithms are proposed to solve NP-complete problems, such as
Belief Propagation and Survey Propagation \cite{sat,sp,mar}.

In this paper, a \emph{mutual-determination} structure is proposed
by statistical mechanic approach to investigate the solution space
of the minimum vertex-cover problem. This structure reflects the
feature of the Hamming distance \cite{cluster,geometry} among
solutions and describes how tight the correlations among unfrozen
variables are. By this structure, we can detect the equivalent
variables in the solution space \cite{its2010}, i.e., the variables
must take the same or the opposite Boolean values. Furthermore,
based on the existence of the mutual-determination in the solution
space of Vertex-cover, the ranks of nodes of a graph by the
leaf-removal process are provided to describe the influence orders
of leaves in different levels. Taking the advantage of the
leaf-removal ranks and the relationship of mutual-determination with
the backbone and unfrozen variables, we can have a much clear
understanding of the evolution of the states in the solution space
when a new node is added, and a \emph{reduced solution graph} is
defined to exactly express the structural information of the
solution (sub-)space. Finally, an algorithm named
\emph{Mutual-determination and Backbone Evolution Algorithm} is
proposed by the evolution of the mutual-determinations and backbones
on the reduced solution graph, and some analysis and numerical
experiments are given to verify its efficiency and adaptability.
This algorithm is complete to find the whole solution space of
Vertex-cover when there is no leaf-removal core in the graph,
otherwise an approximated one with relatively better efficiency than
the replica symmetry methods.


\section{Definition of interaction}

A vertex cover on an undirected graph $G(N,M)$ with $N$ nodes and
$M$ edges is a subset $S=\{i_1, i_2, \cdots, i_m\}$ of its nodes
such that every edge has at least one endpoint in $S$. The minimum
vertex-cover problem is an optimization problem to find the minimum
size of a vertex cover on a given graph. Mapped to spin-glass model,
energy function of the minimum vertex-cover problem can be written
as
\begin{equation}
E[\{\sigma_i\}]=-\sum_{i=1}^N \sigma_i+\sum_{(i,j)\in
E(G)}(1+\sigma_i)(1+\sigma_j),
\end{equation}
where $E(G)$ denotes the edge set and $(i,j)$s are edges in it,
spin/variable $\sigma_i=-1$ if node $i\in S$ (covered) and
$\sigma_i=1$ otherwise. Then, different energy levels are produced
by different assignments or \emph{configurations} in terminology of
spin-glass theory. The assignments with the lowest energy are named
\emph{solutions/ground states}, and the set of all these solutions
achieving the lowest energy (minimum vertex cover) is named
\emph{solution space} $\mathcal {S}$.

Backbones \cite{backbone} and long-range correlations
\cite{zhou,PNAS} are both the typical structures of solution space
of combinatorial optimization problems, which have been well studied
in algorithmic and statistical analysis. In the solution space
$\mathcal {S}$, spin $\sigma_i$ is \emph{frozen} or called
\emph{backbone} if it takes the same value in all solutions;
otherwise it is \emph{unfrozen}. For an unfrozen spin $\sigma_i$, if
it taking some value will make influence on infinite number of other
spins (assumed $\mathcal {O}(N)$ with the total number of  $N$
spins), it belongs to \emph{long-range correlation}
\cite{frus,PNAS}. Recent research suggests that the complicated
organizations of the solutions of combinatorial optimizations, e.g.,
backbones and long-range correlations, would be the kernel reason
for the algorithmic hardness to find a solution for large-scale
combinatorial optimization problems with massive constraints and
variables \cite{PHASE}. To study the solution space $\mathcal {S}$
of Vertex-cover, we classify the variables as unfrozen,
\emph{positively frozen} (frozen to $+1$) and \emph{negatively
frozen} (frozen to $-1$) variables.

As a generalization of the backbone and long-range correlation, a
new structure named \emph{mutual-determination} is proposed to
achieve a better understanding of the solution space, which can be
viewed as an interactive relation of unfrozen variables in the
solution space. If two unfrozen variables form a
mutual-determination, the fixation of the assignment of any one will
result in the fixation of the other in the solution space. Indeed,
it is a special relation implied by the constraints that two
unfrozen variables can be mutually determined by each other, i.e.,
if two unfrozen variables $\sigma_i, \sigma_j$ form a
mutual-determination, then $\sigma_i+\sigma_j=0$. When two unfrozen
nodes form mutual-determination for Vertex-cover, it means that if
one of them is covered, the other should be uncovered, and there is
one and only one should be covered for this pair of nodes.

By the famous survey propagation algorithm \cite{sp}, it takes
advantage of the idea of the backbone and long-range correlation to
gradually eliminate variables and constraints of the original
problem in size, and achieves excellent performance for solving
3-SAT, Vertex-cover, etc. As the motivation for proposing the
mutual-determination structure and for that the variables in
interactive relations are equivalent variables, we can use a simple
logic $\sigma_j=-\sigma_i$ to decrease the number of variables in
the original problem to obtain new algorithmic strategies. In the
following sections, we will use the backbones and
mutual-determinations to analyze the solution space of Vertex-cover.

\section{Reduced solution graph of Vertex-cover}

To study the solution space of Vertex-cover, the leaf removal
\cite{leafremoval} should be mentioned as inspiration. Given a graph
$G$, a \emph{leaf} is a couple of nodes $\{v,w\}$ in which the first
one has degree 1 and the second one is the only neighbor of it.
Here, node $v$ is a pendant point in the graph, node $w$ acts as a
petiole, and for the same petiole there may be more than one pendant
points connecting it. To define the \emph{leaf removal}, if the
nodes pair $\{v,w\}$ is a leaf in graph $G$, remove the two nodes
with the edges touching them. It is very interesting that the leaf
removal process can destroy all the leaves in graph $G$ and can
produce new leaves for the rest graph. In Figure 1, a leaf removal
process for a simple graph is shown.

\begin{figure}[!t]
\centering
\includegraphics[width=5in]{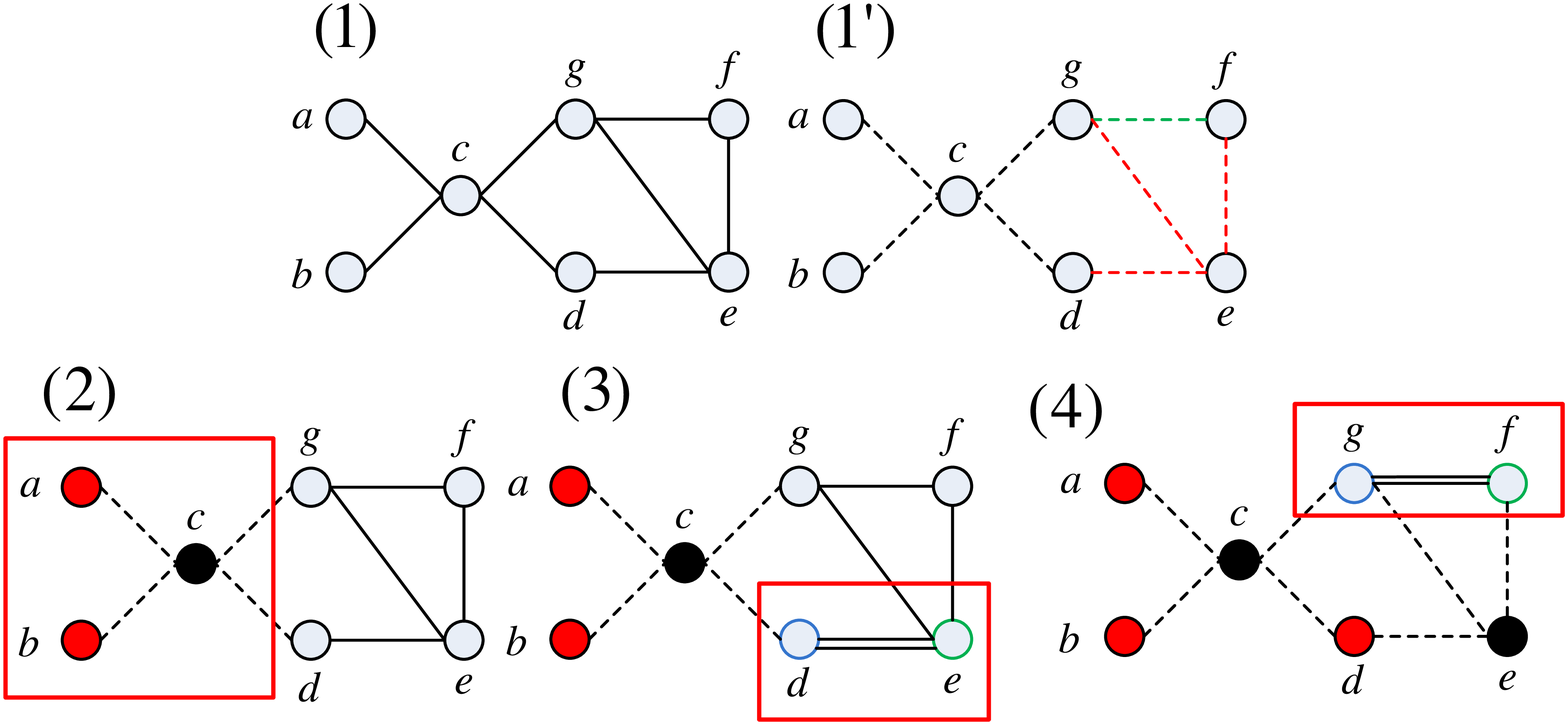}
\caption{The subgraph (1) is the original graph and the subgraph
(1') reflects the leaf removal process. For the original graph (1),
nodes $\{a,b,c\}$ form a multiple leaf with the common petiole $c$
and the corresponding edges are marked by black dashed lines; after
removing this leaf, a new leaf with nodes $\{d,e\}$ appears and the
corresponding edges are marked by red dashed lines; when leaf
$\{d,e\}$ is removed, the last new leaf $\{f,g\}$ is produced and
the corresponding edges are marked by green dashed lines. The
subgraphs (2-4) reveal the relationship of nodes in the leaves of a
graph. There are three leaves in the graph which are marked by the
red rectangles in subgraph (2-4), and the relations among the nodes
are revealed by the constraints in the leaves for Vertex-cover.}
\label{fig1}
\end{figure}

By this leaf removal process, we can find that there exist some
graphs which have no leaf removal core until the termination of this
process, which means that each node belongs to a leaf in the graph.
For the Vertex-cover of these graphs, to obtain the minimum vertex
cover for the graph, there is one and only one node should be
covered for each leaf. By the results in \cite{leafremoval}, a
trivial minimum vertex cover can be obtained by making all the
petioles covered with all the pendant points uncovered in different
levels of leaves, e.g., making the nodes $\{c,e,g\}$ or $\{c,e,f\}$
covered and the rest uncovered leads to minimum vertex covers. Here,
we will take use of this trivial solution to construct a
relationship between/among nodes in a leaf.

In Figure 1, the first subgraph (1) is the original graph. For the
first leaf $\{a,b,c\}$ in this graph which lies in the red rectangle
of subgraph (2), there are two pendant points with one petiole, and
to ensure the minimum coverage of the subgraph of $\{a,b,c\}$, the
only way is to cover the petiole node $c$ and make the two pendant
points $\{a,b\}$ uncovered. In this case, the node $c$ acts as a
negatively frozen node (backbone), which is marked by solid black
circles, the node $a,b$ acts as a positively frozen node, which is
marked by solid red circles, and the edges connecting them are
marked by dashed ones. For the second leaf $\{d,e\}$ in the red
rectangle of subgraph (3), there is only one pendant point $d$ and
one petiole $e$, to ensure the minimum coverage of the subgraph of
$\{a,b,c,d,e\}$, covering any node of $\{d,e\}$ with the other one
uncovered will make an optimization solution. In this case, the
assignments of $\sigma_d,\sigma_e$ must be opposite and they are
\emph{mutually determined} in the solution space of Vertex-cover,
which is denoted by a double edge and two nodes with different
colors. For the third leaf $\{g,f\}$ which is similar as the leaf
$\{d,e\}$, their relation is also mutually determination. However,
the relation of $\{g,f\}$ makes influence on the leaf $\{d,e\}$,
which makes the states of nodes $\{d,e\}$ changed to be backbones,
and the detailed techniques for this case will be discussed in the
following sections.

Based on the analysis inspired by the example in Figure 1 and the
backbone and mutual-determination structures, we can construct an
expression of the solution space of Vertex-cover which named
\emph{reduced solution graph} $\mathcal {S}(G)$: to show different
minimum vertex covers of a given graph $G$, the backbones on it are
marked by solid red or black circles, and \emph{double edges}
between unfilled hollow nodes with different colors (blue and green)
suggest that the relations between the nodes are
mutual-determinations and they can not take the same value
simultaneously; the edges connecting the backbones will be changed
to dashed ones and the edges connecting two unfrozen nodes are
retained. By the leaf-removal process and the strong correlation
among/between nodes in the leaves, the mutual-determination can only
be in the pendant point and its petiole. Then, it is evident that
the reduced solution graph can express the solution space of
Vertex-cover strictly when there is no leaf removal core for the
given graph, and whether this expression is also effective for
general graphs with leaf-removal cores and how to obtain the reduced
solution graph of Vertex-cover will be discussed in the following.

In order to have a convenient analysis of the reduced solution
graph, the leaf-removal \cite{leafremoval} sequence is very
important. Here, we take advantage of the sequence of the leaf
removal to define the rank of each node:

\emph{Step }1: All the pendant points in the graph are assigned to
sequence order/rank 1, and their neighbors (the petioles) have rank
2;

\emph{Step }2: Remove the leaves with edges connecting them from the
graph. After the leaf removal, all the new produced pendant points
are assigned to rank 3, and their corresponding petioles have rank
4;

\emph{Step }3: Repeat the steps 1-2 and assign increasing ranks
until their are no new leaves produced. If there is still a
leaf-removal core after the above two steps, assign the nodes in the
core with ranks according to their already ranked neighbors by
gradually increase.

\section{Analysis of mutual-determination in the solution space of Vertex-cover}

In this section, we concern on achieving the reduced solution graph
$\mathcal {S}(G)$ by determining the states of the nodes one by one
following the leaf removal sequence/ranks. This process is fulfilled
by a method similar to the cavity method, and for each node its
state is determined by the local environment of itself. Considering
a new node $i$ connected to a graph $G$ with $k$ edges, the newly
produced graph is denoted by $G'$. For the neighborhood of node $i$,
there are three kinds of neighbors: positively frozen ones,
negatively frozen ones and unfrozen ones.

\subsection{Local evolution of mutual-determinations and backbones}

In this subsection, we consider the different local environments of
a new node $i$, and investigate the state determination and
evolution of it and associated nodes. Taking advantage of the
analysis in \cite{frus}, we first study the following four cases:

$\bullet$ \textbf{\emph{Case} A}: only one of its neighbors is
positively frozen in $G$; some other neighbors are unfrozen nodes
which can take spin value $-1$ simultaneously.

In case A, energy increase is unavoidable when node $i$ is added.
When $\sigma_i$ takes value $-1$ (covered), its neighbors of nodes
are free to take their spin values in the original $G$, and new
covers of the new graph $G'$ come out with the lowest energy; when
$\sigma_i$ takes value $1$, the positively frozen neighbor should be
changed to an unfrozen node taking $-1$ now, e.g., adding the node
$e$ to the subgraph of $\{a,b,c,d\}$ leads a mutual-determination of
$\{d,e\}$ in Figure 1 and the above subgraph in Figure 3 also show
this process. Then, mutual-determination of the new added $i$ with
the original positively frozen node $j$ is formed.

 $\bullet$ \textbf{\emph{Case}
B}: there are more than two neighbors which are positively frozen.

In case B, energy increase is unavoidable when node $i$ is added. To
obtain a coverage of the new graph $G'$ from the original one $G$,
the new added node $i$ must be covered without other choice. Then,
node $i$ is negatively frozen. (There will be a supplementary and
additional adjustment for this case in the following case E when the
positively frozen neighbors have some common properties.)

$\bullet$ \textbf{\emph{Case} C}: there is no neighbor which is
positively frozen, but all the unfrozen neighbors can take spin
value $-1$ simultaneously.

In case C, energy increase is avoidable when node $i$ is added, and
the new added node $i$ should be uncovered. When $\sigma_i$ takes
value $+1$, all the unfrozen neighbors should take $-1$
simultaneously. Then, by the mutual-determinations and coverage of
each edge in the reduced solution graph $\mathcal {S}(G)$, that
these unfrozen neighbors change to be negatively frozen will lead a
number of associated unfrozen nodes to be frozen.

$\bullet$ \textbf{\emph{Case }D}: there is at least one pair of
unfrozen neighbors that can not take spin value $-1$ simultaneously.

In case D, as the two neighbors can not take $-1$ simultaneously,
energy increase is inevitable. Then, the new added variable $i$
should take $-1$ to ensure the coverage. However, for the
incompatible cycles (like that in the below subgraph of Figure 3),
making any other node except $i$ frozen to $-1$ and the rest
unfrozen nodes connected by alternatively existing double edges will
have the same effect and ensure the coverage. Thus, in case D
freezing the new added node $i$ to $-1$ will reduce the whole
solution space to a partial solution subspace. Nevertheless, as
takeing any one to be frozen on the incompatible cycle leads to a
solution subspace with the same size, we have the convenient way to
make $i$ negative backbone.

By the above analysis, the incompatible cycle in the reduced
solution graph makes a possible inaccurate choice of the negatively
frozen backbone. Thus, the hardness of solving Vertex-cover mainly
stems from the incompatible cycles.

There exists an interesting entanglement between case A and case C.
In case C, some added node $i_1$ has a positively frozen state, its
unfrozen neighbors are forced to be frozen with some associated
nodes. In case A, if some new added node $i_2$ is connected to $i_1$
and forms mutual-determination with $i_1$ by the rule of case A, the
nodes that have been frozen by $i_1$ should be released to their
original unfrozen state. To fulfill this releasing steps, an
additional mark should be sticken to the node number, e.g., a node
$(4, 7)$ means that the node 4 has been frozen by the operation of
adding node 7. Indeed, this \emph{freezing influence} happens only
in case C with node 7 positively frozen. Then, if the state of node
7 is changed to be unfrozen by adding a new node 8 in case A, we can
release all the nodes with mark $(*, 7)$ and change the
corresponding numbers to $(*, 0)$, in which 0 means the state of the
node is unfrozen. This operation is named \emph{releasing
operation}.

For the releasing operation, there is a special case should be
considered for the node adding process:

$\bullet$ \textbf{\emph{Case }E}: there are more than two neighbors
which are positively frozen and have the same additional mark, and
the unfrozen neighbors can take spin value $-1$ simultaneously. This
case is a supplementary and additional adjustment of case B.

In this case, the current node should form mutual-determinations
with the positively frozen nodes whose additional marks are the
same, and the releasing operation is operated for these positively
frozen neighbors (a simple example can be seen in the below subgraph
of Figure 7).

\subsection{Some supplementing techniques for the states evolution}

\begin{figure}[!t]
\centering
\includegraphics[width=5in]{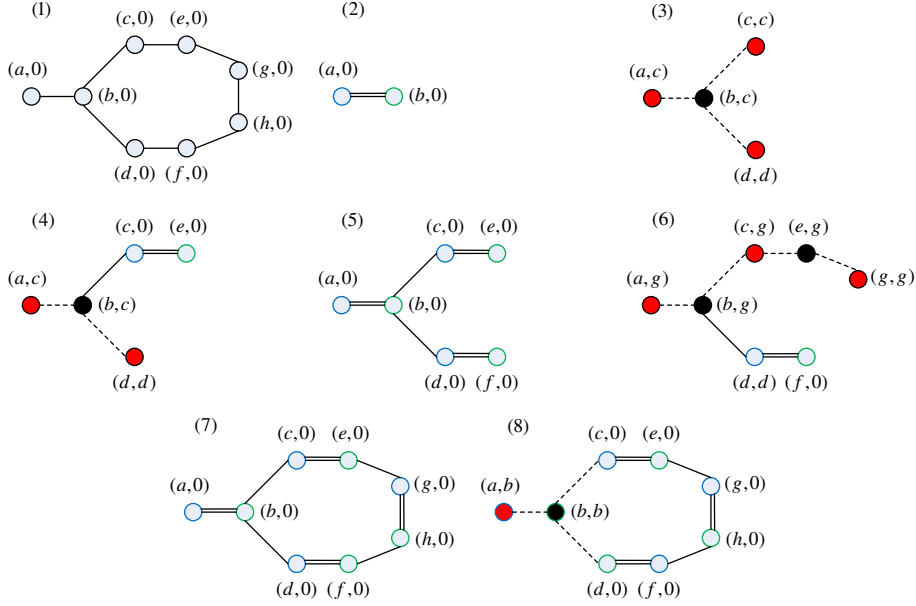}
\caption{An example for the emergence of odd cycles of unfrozen
nodes and the way to break this conflicted cycle. Subgraph (1)
provides the original graph for Vertex-cover; subgraph (2) describes
the process of adding the nodes $\{a,b\}$, which consists a
procedure of case A to produce a mutual-determination; subgraph (3)
describes the process of adding the nodes $\{c,d\}$, which consists
a procedure of case C to produce a positively frozen backbone;
subgraph (4-5) is for adding $e$ and $f$ separately, which consists
a procedure of case A again, and the checking technique works when
adding $e$ and the rechecking technique works when adding $f$;
subgraphs (6-7) correspond the process of adding $g,h$, which is
obtained by case A and C, and the freezing influence work when
adding $g$ and the releasing operation works when adding $h$;
subgraph (8) is obtained by breaking the odd cycle to obtain the
real reduced solution graph, which changes the node $b$ with lowest
rank to be negatively frozen.} \label{fig2}
\end{figure}

In the releasing operation for case A, to avoid some possible
mistakes, a \emph{checking technique} should be considered. When
releasing the negatively frozen backbones in the releasing
operations for case A, its local environment should be considered,
and if there are positively frozen neighbors for the current
negatively frozen backbone whose additional mark is not same as
itself, the releasing process should be stopped (e.g., the node $b$
in the process of subgraphs (3-4) in Figure 2).

Be specific to this checking item, after the operations of case A-D,
a \emph{rechecking technique} should be added: when the freezing
influence and releasing operations of adding a new node have been
done, we should check any of the negatively frozen backbone whose
additional mark is not 0, if there is only one positively frozen
neighbor for itself, release the negatively frozen backbone with the
only positively frozen neighbor and the nodes which have the same
additional mark with it (e.g., the node $b$ in the process of
subgraphs (4-5) in Figure 2).

At last, by the process of above analysis, a complicated
structure-odd cycles on the reduced solution graph could come into
appearance, which makes conflictions for the relations among the
unfrozen nodes. For the example in Figure 2, the subgraphs (1-7)
provide the process from the original graph to the reduced solution
graph by adding the nodes one by one using our techniques above, but
unfortunately the unfrozen nodes $b,c,d,e,f,g,h$ in subgraph (7)
form an odd cycle. In the odd cycle, we find that any node except
$b$ taking any value will force the node $b$ to be negatively
frozen, and it is an incompatible cycle. To break this disharmony,
the only way is to change the state of $b$ to be negative backbone
and make corresponding changes for its neighbors with lower ranks
(e.g., the process from subgraph (7) to (8) in Figure 2). This
technique is named \emph{odd cycles breaking}.

\subsection{Global characteristics of mutual-determinations}

In the following, we will have an explicit discussion of the
mutual-determinations and unfrozen nodes structure of Vertex-cover.
As mentioned above, if some node $i$ forms mutual-determination with
a node $j$, i.e., $\sigma_i=-1$ forces $\sigma_j$ to take $+1$,
correspondingly by the Vertex-cover, we have that $\sigma_i=+1$
requires $\sigma_j=-1$ to satisfy the coverage. If a node $j$ forms
a mutual-determination chain with some other nodes
$j_0,j_1,\cdots,j_k$, a possible way is that the edges
$(j,j_0),(j_1,j_2),\cdots$ on the reduced solution graph are all
double edges, i.e., all these pairs of nodes form
mutual-determinations, which is shown in the Cycle 2-Compatible
Cycle in Figure 3. When the node $j$ takes vale $-1$ (covered), the
nodes $j_0, j_2, \cdots, j_{2l},\cdots$ must take $+1$ by the
mutual-determination relations and coverage of the edges connecting
them, and the nodes $j_1, j_3, \cdots, j_{2l+1},\cdots$ must take
$-1$. Therefore, the alternatively existing double edges on the
reduced solution graph lead to the emergence of the strong
correlation for nodes of long distance.

\begin{figure}[!t]
\centering
\includegraphics[width=5in]{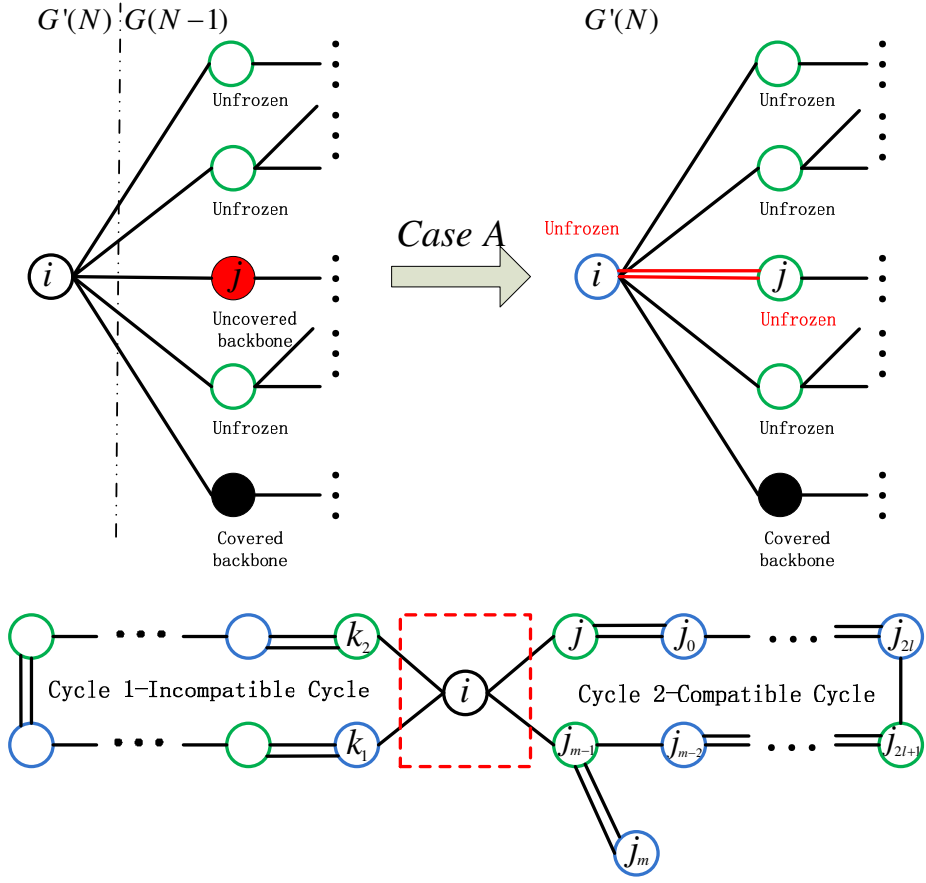}
\caption{The above subgraph provides the process of case A to
produce the mutual-determination structure, and the below subgraph
reveals the formation of incompatible cycle and compatible cycle by
interaction.} \label{fig3}
\end{figure}

Without confusion, we neglect the backbones in the reduced solution
graph but keep the unfrozen nodes. As we know, there are almost no
local cycles on random graphs and the cycle sizes on random graph
are of $O(log N)$. By this characteristic of random graph, the
emergence of the Cycle 2-Compatible Cycle in Figure 3 leads to the
long-range correlation structure, and indeed the alternating
mutual-determination chain is the only way to produce the long-range
correlation in Vertex-cover.

When the unfrozen neighbors of the new added node $i$ only have
influence range over tree structures, which means that the double
edges belonging to different unfrozen neighbors are disconnected
except $i$. At this time, these unfrozen neighbors can not propagate
its influence to each other, and they can take $-1$ simultaneously.
Similar as that in random graph, as the increasing of the number of
nodes and edges, the unfrozen nodes with double edges in the reduced
solution graph may connect together, and form cycles and even giant
connected component. Especially, for random graphs, the cycles
connected by unfrozen nodes in the reduced solution graph must be
with size of $O(logN)$. Therefore, if the unfrozen nodes connect
together to form a giant connected component \cite{giant}, some of
them taking value $-1$ will cause a percolation phenomenon
\cite{network} that many other nodes ($O(N)$) in this giant
connected component will be forced to be frozen. As a result, the
long-range correlation phenomenon emerges. In the work of Zhou
\cite{frus}, the long-range correlation of Vertex-cover for random
graph appears at $c=e$. Indeed, by the literature of statistical
mechanics, the existence of long-range correlation has close
connection with the replica symmetric breaking of the solution
space. As the correlation is formed by mutual-determinations, the
long-range correlation can also provide an explicit explanation of
the clustering phenomenon of solution space.

By the emergency of the long-range correlation nodes, the local
environment becomes much more complicated. As unfrozen neighbors of
the new added node $i$ can be connected together by other unfrozen
nodes, their values can not be assigned arbitrarily. In Figure 3, a
schematic view of the compatible and incompatible cycles of unfrozen
nodes is shown. In Cycle 1, the unfrozen nodes $k_1,k_2$ are
connected by unfrozen nodes with alternatively existing double
edges, and it is easy to find that the nodes $k_1,k_2$ form a
long-range correlation relation and can not take spin value $-1$
simultaneously. To the contrary, in the compatible Cycle 2, though
the relation between $j,j_{m-1}$ is also long-range correlation,
they can take spin value $-1$ simultaneously by the
mutual-determination chain in Cycle 2.

\section{Mutual-determination and Backbone
Evolution Algorithm for Vertex-cover}

In this section, we will introduce an algorithm for solving
Vertex-cover based on case A-E and the node ranks. By the analysis
in case A, B, C, D, E, we consider to update the state of the
original graph $G(N-1)$ after adding a new node $i$. As the node
states are classified by mutual-determinations and backbones, we can
get an algorithm to find the reduced solution graph $R(G)$ of
Vertex-cover, this algorithm is named \emph{Mutual-determination and
Backbone Evolution
Algorithm}, shown as follows:\\

Let's take $G[v]$ as an induced subgraph of original graph $G$ by
adding node $v$ and $E_{ij}=0(1)$ represent nodes $i$ and $j$ are
connected (unconnected). Especially, when nodes $i$ and $j$ form the
mutual-determination, we take $E_{ij}=2$. $backbone[i]=-1,0$ or $1$
means the node $i$ is unfrozen, positively frozen or negatively
frozen node respectively; $root[i]$ is the additional mark of node
$i$.\\

\quad \\
The algorithm of \emph{Releasing Operation} is shown as follows:
\quad \\

\noindent
\begin{tabular}[c]{p{480pt}l}
\textbf{algorithm}  \textbf{Releasing($i$,$G$,$ROOT$)} \\
\textbf{begin}\\
\ \ \ \ \ \ \ \ \texttt{$backbone[i]=-1$;}\\
\ \ \ \ \ \ \ \ \texttt{$root[i]=0$;}\\
\ \ \ \ \ \ \ \ \textbf{for} \ \texttt{all neighbor $j$ of $i$ in $G$ with $root[j]=ROOT$} \textbf{do}\\
\ \ \ \ \ \ \ \ \textbf{begin}\\
\ \ \ \ \ \ \ \ \ \ \ \ \ \ \ \ \ \textbf{if}(\texttt{$backbone[j]=1$ and $j$ has at least one positively frozen}\\
\ \ \ \ \ \ \ \ \ \ \ \ \ \ \ \ \ \texttt{neighbor $k$ with $root[k]\neq ROOT$})\ \textbf{then}\\
\ \ \ \ \ \ \ \ \ \ \ \ \ \ \ \ \ \ \ \ \ \ \ \ \ \ \texttt{continue;}\\
\ \ \ \ \ \ \ \ \ \ \ \ \ \ \ \ \ \textbf{else} \textbf{do}\\
\ \ \ \ \ \ \ \ \ \ \ \ \ \ \ \ \ \ \ \ \ \ \ \ \ \ \textbf{Releasing($j$,$G$,$ROOT$);}\\
\ \ \ \ \ \ \ \ \textbf{end}\\
\textbf{end}\\
\end{tabular}

\quad \\
The algorithm of \emph{Freezing Influence} is shown as follows:
\quad \\

\noindent
\begin{tabular}[c]{p{480pt}l}
\textbf{algorithm}  \textbf{Freezing($i$,$G$)} \\
\textbf{begin}\\
\ \ \ \ \ \ \ \ \textbf{if}(\texttt{$backbone[i]=0$}) \ \textbf{do}\\
\ \ \ \ \ \ \ \ \ \ \ \ \ \ \ \ \ \textbf{for} \texttt{all unfrozen neighbor $j$ of $i$ in $G$} \textbf{do}\\
\ \ \ \ \ \ \ \ \ \ \ \ \ \ \ \ \ \ \ \ \ \ \ \ \ \ \texttt{$root[j]=root[i]$;}\\
\ \ \ \ \ \ \ \ \ \ \ \ \ \ \ \ \ \ \ \ \ \ \ \ \ \ \texttt{$backbone[j]=1$;}\\
\ \ \ \ \ \ \ \ \ \ \ \ \ \ \ \ \ \ \ \ \ \ \ \ \ \ \textbf{Freezing($j$,$G$);}\\
\ \ \ \ \ \ \ \ \textbf{if}(\texttt{$backbone[i]=1$}) \ \textbf{do}\\
\ \ \ \ \ \ \ \ \ \ \ \ \ \ \ \ \ \textbf{for} \texttt{all unfrozen neighbor $j$ of $i$ in $G$ with $E_{ij}=2$} \textbf{do}\\
\ \ \ \ \ \ \ \ \ \ \ \ \ \ \ \ \ \ \ \ \ \ \ \ \ \ \texttt{$root[j]=root[i]$;}\\
\ \ \ \ \ \ \ \ \ \ \ \ \ \ \ \ \ \ \ \ \ \ \ \ \ \ \texttt{$backbone[j]=0$;}\\
\ \ \ \ \ \ \ \ \ \ \ \ \ \ \ \ \ \ \ \ \ \ \ \ \ \ \textbf{Freezing($j$,$G$);}\\
\textbf{end}\\
\end{tabular}

\quad \\
Now, it is the algorithm of Mutual-determination and Backbone
Evolution Algorithm:
\quad \\

\noindent
\begin{tabular}[c]{p{480pt}l}
\textbf{begin}\\
\ \ \ \ \texttt{calculate the leaf-removal sequence $layer[i]$ of $G$;}\\
\ \ \ \ \texttt{vertex set $v=\emptyset$;}\\
\ \ \ \ \texttt{initialize $backbone[i]=-1$ and $root[i]=0$ for all $i$ in $G$;}\\
\ \ \ \ \textbf{for} \texttt{$l=1$ to Max($layer[i]$,$1\leq i\leq N$)} \textbf{do}\\
\ \ \ \ \textbf{begin}\\
\ \ \ \ \ \ \ \ \textbf{for} \texttt{all vertex $i$ with $layer[i]=l$} \textbf{do}\\
\ \ \ \ \ \ \ \ \textbf{begin}\\
\ \ \ \ \ \ \ \ \ \ \ \ \texttt{$v=v\cup i$;}\\
\ \ \ \ \ \ \ \ \ \ \ \ \texttt{$G'=G[v]$;}\\
\ \ \ \ \ \ \ \ \ \ \ \ \texttt{calculate $num$ is the number of positively frozen neighbors of $i$ in $G'$}\\
\ \ \ \ \ \ \ \ \ \ \ \ \textbf{if}(\texttt{$num=1$}) \ \textbf{do}\\
\ \ \ \ \ \ \ \ \ \ \ \ \textbf{begin}\\
\ \ \ \ \ \ \ \ \ \ \ \ \ \ \ \ \texttt{$Pos$ is the positively frozen neighbor of $i$;}\\
\ \ \ \ \ \ \ \ \ \ \ \ \ \ \ \ \textbf{if}(\texttt{unfrozen
neighbors of $i$ in $G'$ can take $-1$ simultaneously})\ \textbf{then}\\
\ \ \ \ \ \ \ \ \ \ \ \ \ \ \ \ \ \ \ \ \texttt{$E_{i,Pos}=2$;}\\
\ \ \ \ \ \ \ \ \ \ \ \ \ \ \ \ \ \ \ \ \textbf{Releasing($Pos$,$G'$,$root[Pos]$);}\\
\ \ \ \ \ \ \ \ \ \ \ \ \ \ \ \ \ \ \ \ \textbf{Rechecking technique and Odd cycles breaking;}\hfill *\texttt{\emph{Case} A}*\\
\ \ \ \ \ \ \ \ \ \ \ \ \ \ \ \ \textbf{else} \textbf{do}\\
\ \ \ \ \ \ \ \ \ \ \ \ \ \ \ \ \ \ \ \ \texttt{$backbone[i]=1$;}\\
\ \ \ \ \ \ \ \ \ \ \ \ \ \ \ \ \ \ \ \ \texttt{$root[i]=i$;}\hfill *\texttt{\emph{Case} D}*\\
\ \ \ \ \ \ \ \ \ \ \ \ \textbf{end}\\
\ \ \ \ \ \ \ \ \ \ \ \ \textbf{if}(\texttt{$num\ge 2$}) \ \textbf{do}\\
\ \ \ \ \ \ \ \ \ \ \ \ \textbf{begin}\\
\ \ \ \ \ \ \ \ \ \ \ \ \ \ \ \ \textbf{if}(\texttt{all frozen
neighbors of $i$
in $G'$ have same additional mark and}\\
\ \ \ \ \ \ \ \ \ \ \ \ \ \ \ \ \texttt{unfrozen neighbors can take $-1$ simultaneously}) \textbf{then}\\
\ \ \ \ \ \ \ \ \ \ \ \ \ \ \ \ \ \ \ \ \texttt{$Pos$ is one of neighbors of $i$ randomly;}\\
\ \ \ \ \ \ \ \ \ \ \ \ \ \ \ \ \ \ \ \ \texttt{$E_{i,Pos}=2$;}\\
\ \ \ \ \ \ \ \ \ \ \ \ \ \ \ \ \ \ \ \ \textbf{for} \texttt{all the neighbors $j$ of $i$ in $G'$} \textbf{do}\\
\ \ \ \ \ \ \ \ \ \ \ \ \ \ \ \ \ \ \ \ \ \ \ \ \textbf{Releasing($j$,$G'$,$root[j]$);}\\
\ \ \ \ \ \ \ \ \ \ \ \ \ \ \ \ \ \ \ \ \ \ \ \ \textbf{Rechecking technique and Odd cycles breaking;}\hfill *\texttt{\emph{Case} E}*\\
\ \ \ \ \ \ \ \ \ \ \ \ \ \ \ \ \textbf{else} \textbf{do}\\
\ \ \ \ \ \ \ \ \ \ \ \ \ \ \ \ \ \ \ \ \texttt{$backbone[i]=1$;}\\
\ \ \ \ \ \ \ \ \ \ \ \ \ \ \ \ \ \ \ \ \texttt{$root[i]=i$;}\hfill *\texttt{\emph{Case} B}*\\
\ \ \ \ \ \ \ \ \ \ \ \ \textbf{end} \\
\end{tabular}

\noindent
\begin{tabular}[c]{p{480pt}l}
\ \ \ \ \ \ \ \ \ \ \ \ \textbf{if}(\texttt{$num=0$}) \ \textbf{do}\\
\ \ \ \ \ \ \ \ \ \ \ \ \textbf{begin}\\
\ \ \ \ \ \ \ \ \ \ \ \ \ \ \ \ \textbf{if}(\texttt{unfrozen
neighbors of $i$ in $G'$ can take $-1$ simultaneously})\ \textbf{then}\\
\ \ \ \ \ \ \ \ \ \ \ \ \ \ \ \ \ \ \ \ \texttt{$backbone[i]=0$;}\\
\ \ \ \ \ \ \ \ \ \ \ \ \ \ \ \ \ \ \ \ \texttt{$root[i]=i$;}\\
\ \ \ \ \ \ \ \ \ \ \ \ \ \ \ \ \ \ \ \ \textbf{Freezing($i$,$G'$);}\hfill *\texttt{\emph{Case} C}*\\
\ \ \ \ \ \ \ \ \ \ \ \ \ \ \ \ \textbf{else} \textbf{do}\\
\ \ \ \ \ \ \ \ \ \ \ \ \ \ \ \ \ \ \ \ \texttt{$backbone[i]=1$;}\\
\ \ \ \ \ \ \ \ \ \ \ \ \ \ \ \ \ \ \ \ \texttt{$root[i]=i$;}\hfill *\texttt{\emph{Case} D}*\\
\ \ \ \ \ \ \ \ \ \ \ \ \textbf{end}\\
\ \ \ \ \ \ \ \ \textbf{end}\\
\ \ \ \ \textbf{end}\\
\textbf{end}\\
\end{tabular}

\subsection{Some numerical results of the Mutual-determination and Backbone Evolution Algorithm}

\begin{figure}[!t]
\centering
\includegraphics[width=4in]{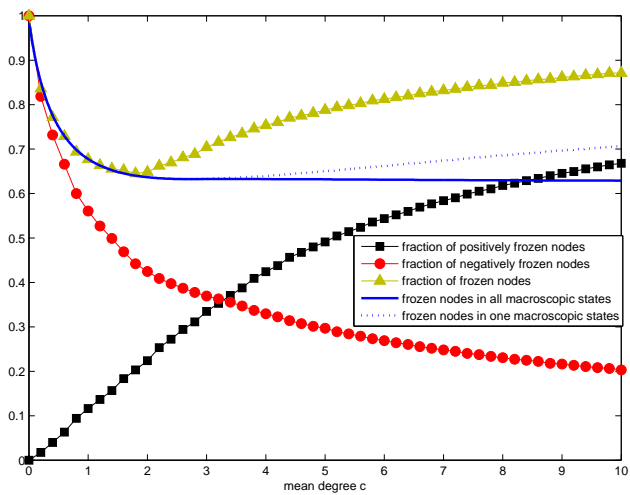}
\caption{Numerical results by our algorithm of fraction of
positively frozen (red solid cycle) and negatively frozen (black
solid rectangle) nodes with different mean degrees, which are
obtained by $1000$ random instances with $N=5000$ nodes; fraction of
frozen vertices in one macroscopic state (yellow solid triangle) and
its comparison with results of Ref.\cite{zhou} (blue line for the
fraction in all macroscopic states and blue dashed line for the
fraction in one macroscopic state). } \label{fig4}
\end{figure}

\begin{figure}[!t]
\centering
\includegraphics[width=4in]{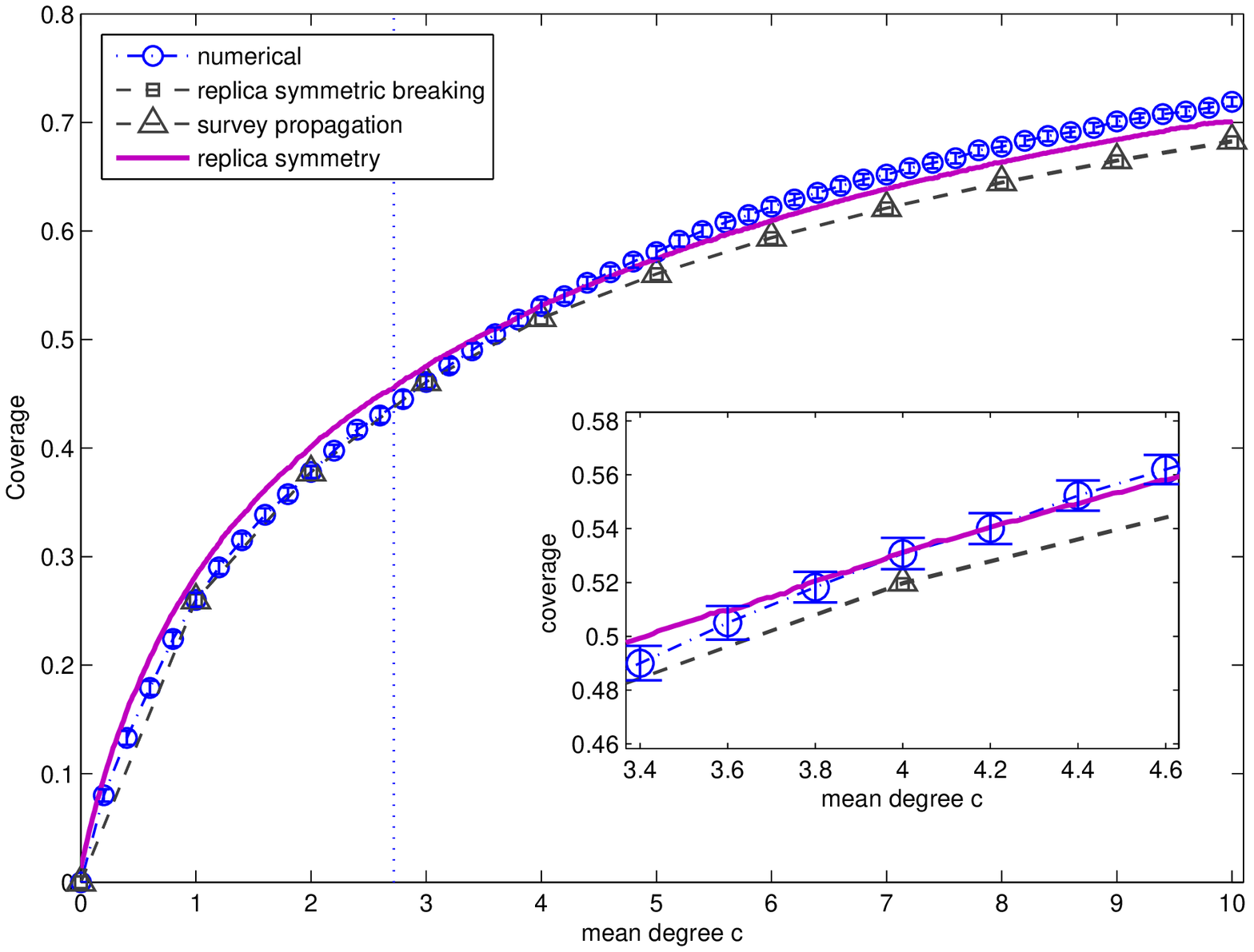}
\caption{Numerical results for Vertex-cover by our algorithm. The
blue cycles with error bar denote the minimal coverage ratio by our
algorithm with $N=5000$ and $1000$ random instances; the brown
rectangles and triangles denote the results of replica symmetric
breaking theory and survey propagation respectively; the purple line
denotes the results by replica symmetry theory; the dashed line
represents mean degree $c=e$.
 }
\label{fig5}
\end{figure}

In this section, some numerical experiments will be performed to
verify the efficiency and performance of the Mutual-determination
and Backbone Evolution Algorithm on random graphs.

To reflect the solution space structures of Vertex-cover, the ratios
of negatively frozen backbone and positively frozen backbones are
detected by the algorithm, in which one is monotone increasing and
the other is monotone decreasing. In Figure 4, the ratios of the
backbones are shown by the solid triangles, and the unfrozen nodes
have its ratio with the residual part of one. Our results on the
frozen nodes in one macroscopic state is higher than that in
\cite{zhou} mainly be the freezing influence. Besides, in Figure 5,
the coverage of the Vertex-cover which is the size of the minimal
vertex-cover is approximated by our algorithm, which is shown by the
blue cycles with error bars and compared with the results of replica
symmetry, replica symmetric breaking theory and survey propagation.
The results on coverage of our algorithm perform better than that of
replica symmetry when the average degree $c$ is not very large, but
still have a small gap with the results of replica symmetric
breaking theory and survey propagation. By the proof of the
strictness of Mutual-determination and Backbone Evolution Algorithm
when $c<e$ in the next section, our numerical results should be
exact ones for the corresponding interval on random graphs.

\begin{figure}[!t]
\centering
\includegraphics[width=5in]{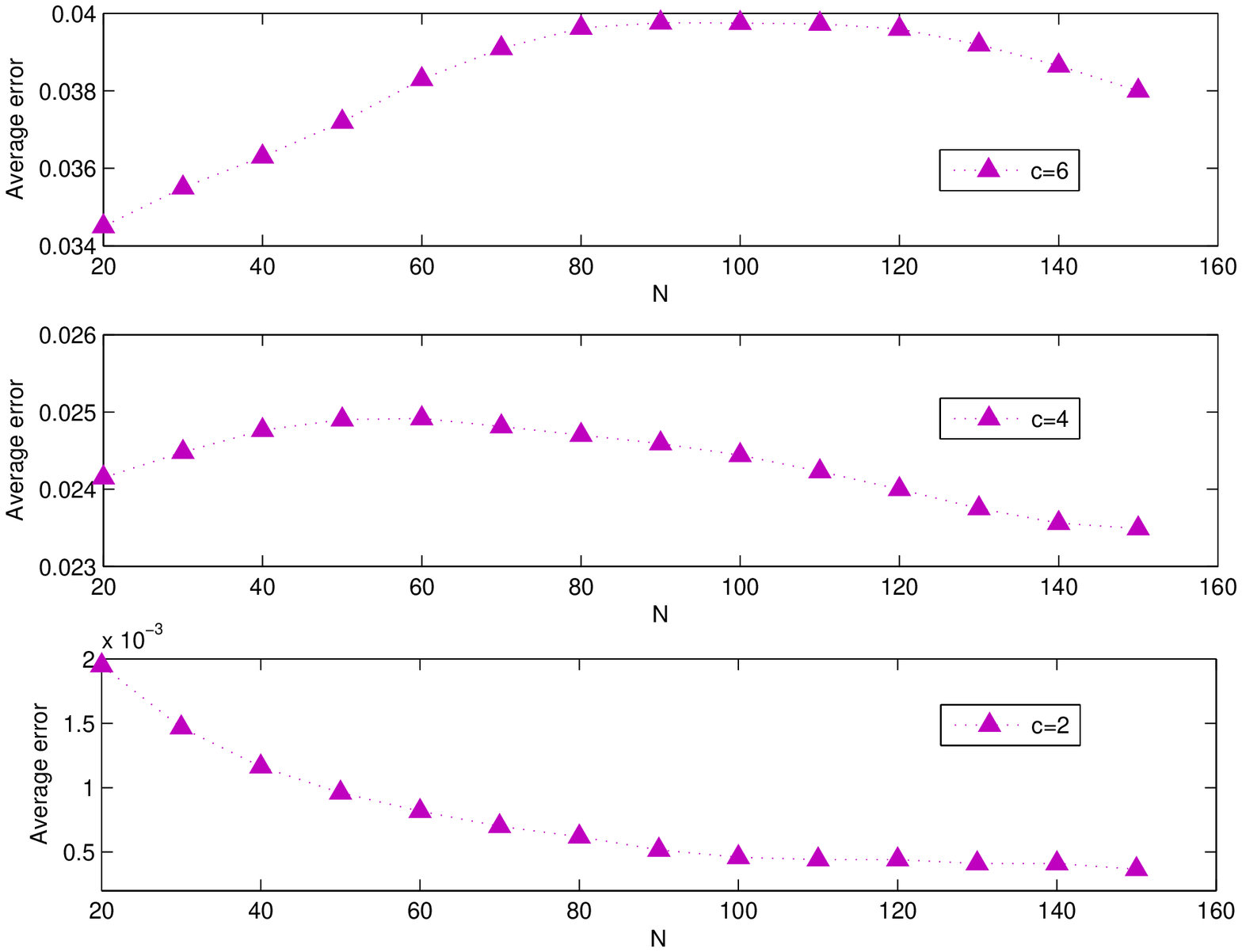}
\caption{Average Error of minimal vertex cover between experimental
and exact results. All results are obtained by $1000$ random
instances with $c=2$, $c=4$ and $c=6$ with different size
$N=20,30,\cdots, 150$. } \label{fig6}
\end{figure}

As a comparison to the complete algorithm and the exact coverage,
some experiments are made to verify the performance of the
Mutual-determination and Backbone Evolution Algorithm which is an
incomplete algorithm. In Figure 6, average Error of minimal vertex
cover between experimental and exact results are plotted to provide
the difference between the exact results and our results on the
coverage, and it is evident to see that these differences are not
very big and have their scales no more than 0.04 for $c=2,4,6$ with
increasing sizes.

\subsection{Mutual-determination and Backbone Evolution Algorithm on some examples}

To detect the reduced solution graph  and provide a primary analysis
of efficiency of our algorithm on the leaf-removal core, we will
discuss the Vertex-cover on the complete graphs and cycles with even
number of nodes for inspiration.

\begin{figure}[!t]
\centering
\includegraphics[width=6in]{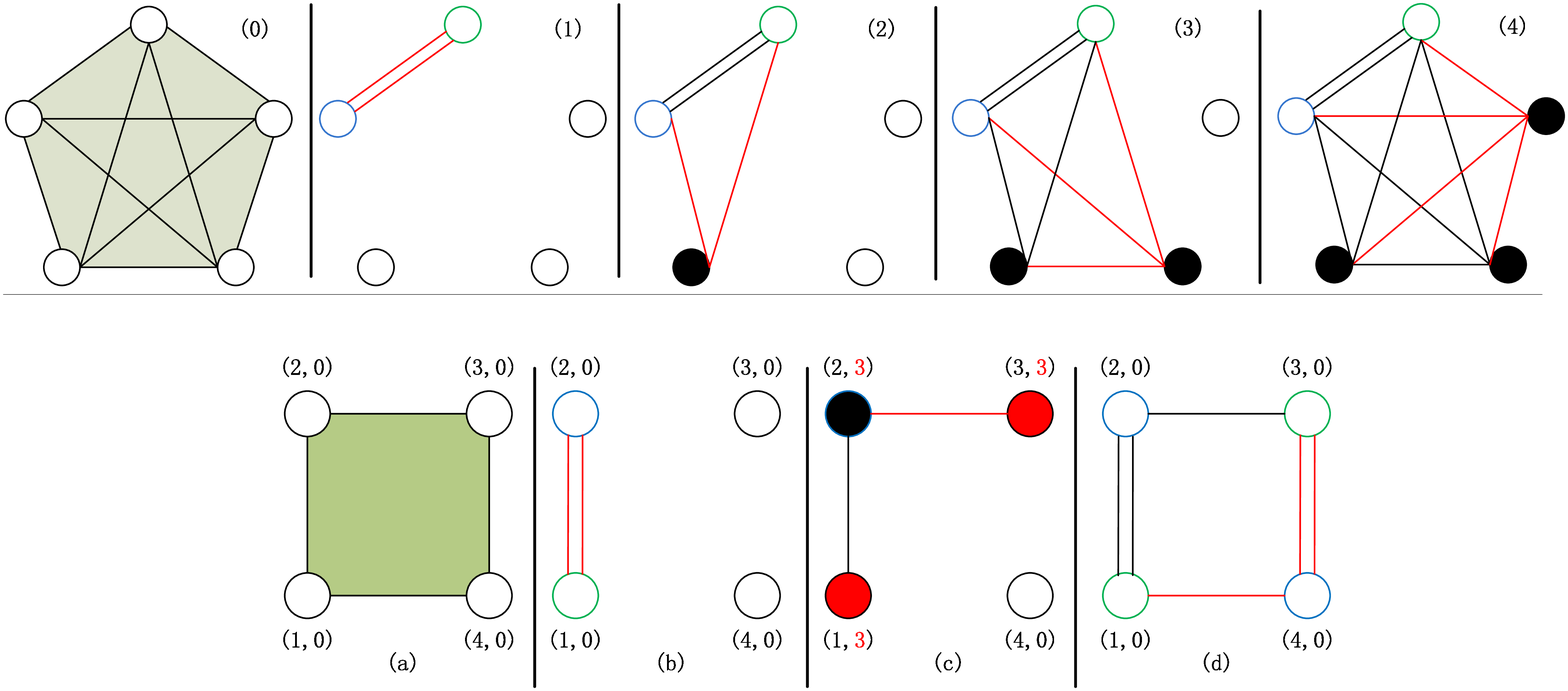}
\caption{The above subgraphs provides the process of obtaining the
reduced solution graph of complete graphs, which is shown by the
complete graph with 5 node in this figure. The below subgraphs
provides the process of obtaining the reduced solution graph of even
cycles.} \label{fig7}
\end{figure}

For the complete graphs, the process of our algorithm to obtain the
reduced solution graph is rather simple: when the second node is
added with an edge (subgraph (1) in Figure 7), the
mutual-determination emerges; for the following added nodes, their
local environment satisfies the case D and they can only be negative
backbones. This process is shown in Figure 7 by a typical graph
$K_5$. It is easy to know that for the complete graph $K_N$ there
must be $N-1$ nodes be covered, and our results of reduced solution
graph correspond to a solution subspace of the Vertex-cover of
$K_N$. The whole solution space possesses $N$ solutions and by our
algorithm we can obtain 2 solutions. Therefore, the
mutual-determination and Backbone Evolution Algorithm is an
incomplete algorithm for the solution space, but it may be efficient
for finding one solution of Vertex-cover. Certainly, as analyzed in
the above section, the incompatible cycles of the unfrozen node will
bring the intrinsic difficulty for solving it and our algorithm can
only obtain some approximated solutions for the original problem.

Then, the cycles $C_{2N}$ with even number of nodes are considered.
The process of our algorithm to obtain its reduced solution graph is
a regular process: when odd number of nodes are added, there are no
unfrozen nodes in the reduced solution graph; when even number of
nodes are added, the releasing operations should be considered and
all the nodes are unfrozen with double edges/mutual-determinations
alternatively connected together; for the last node, it connects
with two positively frozen nodes which have the same additional mark
$2N-1$, and by the case E the last node forms mutual-determination
with node $2N-1$ and the rest are released. The whole process is
schematically shown in the lower subgraphs $(a-d)$ in Figure 7 by a
typical graph $C_4$. Thus, the solution space of $C_{2N}$ can be
obtained, and it is easy to verify that the result is strict by our
algorithm.

\section{Analysis of Mutual-determination and Backbone
Evolution Algorithm for Vertex-cover}

The Mutual-determination and Backbone Evolution Algorithm aims to
obtain the whole solution space, and it is easy to find that it is
an algorithm of polynomial time. As the Vertex-cover problem is a
typical NP-complete problem, this algorithm can not be a complete
one and will lose its efficiency in some case. In this section, some
detailed analysis on the algorithm will be provided.

\subsection{The time complexity of Mutual-determination and Backbone Evolution Algorithm}

By the algorithm in the above section, the process of determining
the ranks of the nodes in the graph is intrinsically a leaf removal
process, and it will cost at most $O(N)$ steps to obtain the whole
ranks of all the nodes.

By considering the nodes sequentially according to their ranks, when
adding a new node to the original graph, first we should consider
its local environment, which will cost at most constant $C$ steps
for random graph. Then, in different cases A-E, there may be
additional time cost. In case B and D, the current state of the new
added node is only determined by its neighbors and it causes no
influence to others; In case C, the freezing of the current node
will cause an influence to the unfrozen neighbors and those related
to them, and thus the influence propagation will cost at most $O(N)$
steps; In case D, the releasing operation with the checking
technique will cost at most $O(N)$ steps. At last, the rechecking
technique and odd cycle breaking will cost at most $O(N)$ steps for
changing the states of some nodes. In sum, when a new node is added
to the graph, there are at most $C+O(N)+O(N)=O(N)$ steps to obtain
the new reduced solution graph.

Besides, by the node ranks, there are $N$ nodes to be added in
total. Therefore, the total time cost for the Mutual-determination
and Backbone Evolution Algorithm is at most $O(N)+N*O(N)=O(N^2)$
steps for random graphs.

\subsection{The strictness of Mutual-determination and Backbone Evolution Algorithm}

\begin{figure}[!t]
\centering
\includegraphics[width=5in]{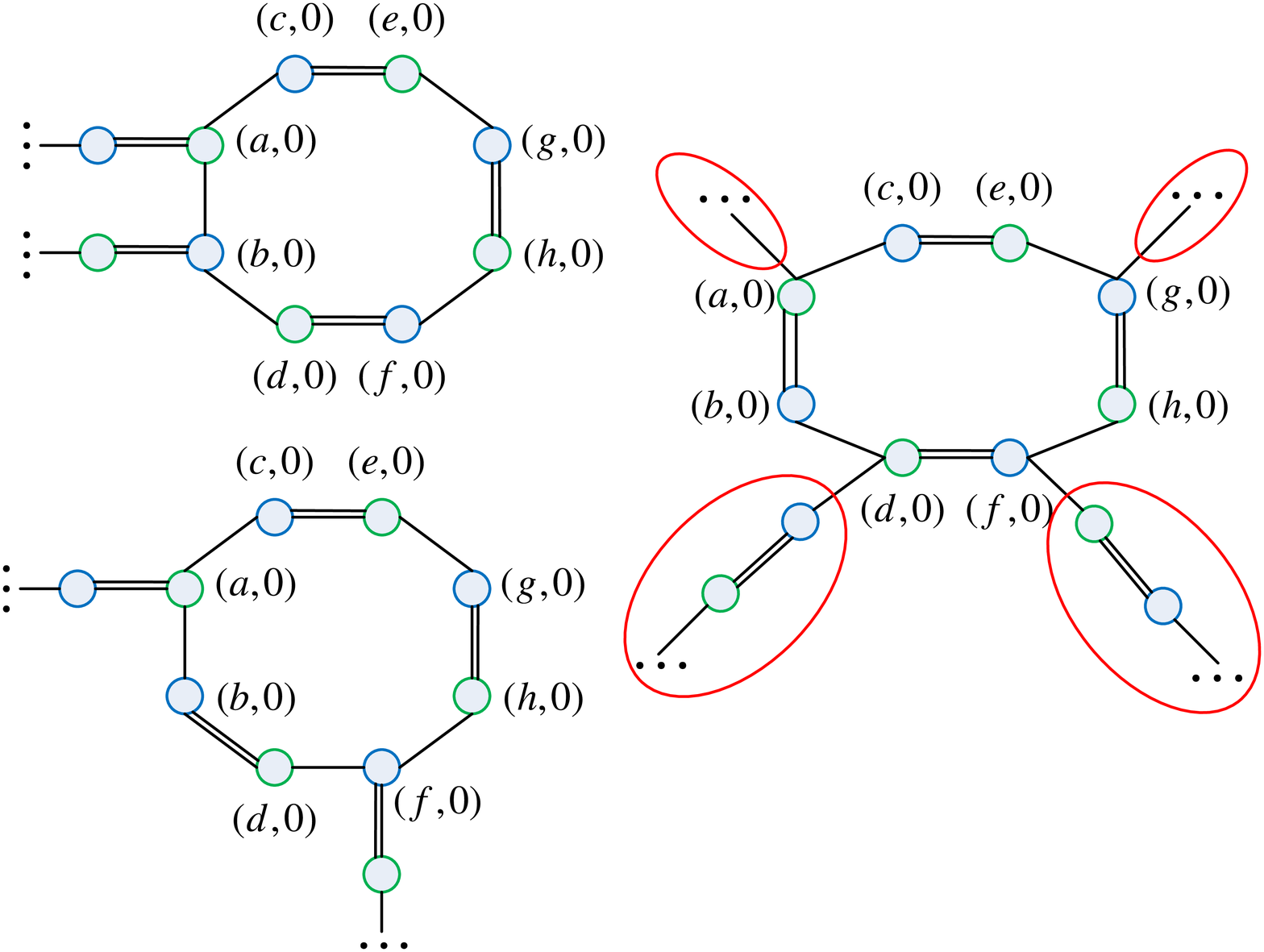}
\caption{Then existence of cycles with non-alternatively
mutual-determinations and the existence of the even cycles with
alternatively mutual-determinations when leaves are removed.}
\label{fig8}
\end{figure}

In this subsection, we will discuss the strictness of the
Mutual-determination and Backbone Evolution Algorithm. By the
analysis in section 3, Vertex-cover can be solved in polynomial time
by assigning the pendants $+1$ and their petioles $-1$ when there is
no leaf removal core. Indeed, if all the nodes can be assigned ranks
by the leaf-removal, i.e., the leaf-removal core is null, the
reduced solution graph can reveal the whole solution space of
Vertex-cover strictly and our algorithm is a complete one to obtain
the whole solution space in this case. The proof is given in the
following.

\noindent\emph{Proof:} For each pair of leaf, they form
mutual-determination or both are backbones with one positive and the
other negative. Our algorithm is intrinsic an evolution process for
the two kinds of states of leaves.

When the reduced solution graph is with unfrozen-node structures of
trees or forests for each step of the algorithm, this evolution
guarantees that each step of adding a leaf will obtain the whole
solution space of the enlarged graph. The strictness of operations
in case B is trivial. Mainly by the case A and C, the releasing
operation and freezing influence are alternatively changing the
states on the trees or forests and have no cross influence among
different branches, which leads to the strictness of our algorithm.

When the reduced solution graph is with unfrozen-node structures of
odd cycles in some steps, the odd cycles breaking technique ensures
the correctness of the algorithm and the resulted reduced solution
graph can be reduced to the case of unfrozen-node structures of
trees or forests above.

The reduced solution graph can never have even cycles with
alternative existing mutual-determinations when there is no
leaf-removal core. In the right subgraph of Figure 8, a schematic
view for the leaf removal is provided. All the nodes in the red
circles will be removed in pairs by leaf removal process, and all
nodes $\{a,b,c,d,e,f,g,h\}$ can only have their own leaf partners on
this cycle, which means that there are no new leaves after the nodes
in the red circles are removed and the even cycle formed by
$\{a,b,c,d,e,f,g,h\}$ survivals at last. Evidently, the even cycles
will be in the leaf-removal core.

Cycles of unfrozen nodes without alternative existing
mutual-determinations can survive on the reduced solution graph. In
the left above and below subgraphs of Figure 8, two simple examples
are given to reveal the existence of ordinary cycles of unfrozen
nodes on the reduced solution graph. By simple logic, we can find
that each node on the graphs can have both covered and uncovered
states. In this situation, the strictness of our algorithm is
guaranteed by the checking and rechecking techniques and the case E,
which ensure that the influence of the freezing operation and
releasing operation can be controlled in a correct way.

At last, considering case D, when the leaf-removal core is null,
this case can be reduced to the odd cycle breaking, and it will
bring the kernel difficulty when the leaf-removal core exists.

Therefore, the reduced solution graph obtained by our algorithm can
reveals the exact solution space when there is no leaf-removal core.
$\Box$

By the above analysis, we have shown that the Mutual-determination
and Backbone Evolution Algorithm is strict when there is no
leaf-removal core in the graph. By the results in
\cite{leafremoval}, there is no leaf-removal core in the random
graph with high probability when the average degree $c$ is less than
$e$. Then, our algorithm is strict with high probability when $c<e$.

\subsection{Cycles in the reduced solution graph }

The even and odd cycles of unfrozen nodes will be analyzed in this
subsection. In this right subgraph of Figure 8, the nodes
$\{a,b,c,d,e,f,g,h\}$ with four mutual-determination
$(a,b),(c,e),(d,f),(g,h)$ construct an even cycle of unfrozen nodes.
Indeed on this cycle all the nodes have a mutual-determination
relation, that is to say, that any node is covered of uncovered will
lead to the fixation of all the other 7 nodes. Then, the double
edges can also be drawn on $(a,c),(b,d),(f,h),(e,g)$ or all the
edges, all these expressions on the reduced solution graph
correspond to the same solution space and there are only two
solutions on the even cycles of alternative mutual-determinations.
For the odd cycles of alternative mutual-determinations, e.g.,
subgraph (7) of Figure 2, what we can do is to perform the odd cycle
breaking, which keeps the strictness of our algorithm.

Unfortunately, there is the other way to produce an odd cycle of
alternative mutual-determinations, just like the Incompatible Cycle
in the below subgraph of Figure 3. This kind of odd cycle structure
emerges when the leaf-removal core exists and is hard to be broken
for the lowest energy configuration of Vertex-cover. As the
incompatible cycle brings obstacle for obtaining the real reduced
solution graph, changing any unfrozen node on it to be negatively
frozen is a possible choice for the reduced solution graph. In our
algorithm, we can only choose one way to proceed, which makes the
solution space collapse to a subspace. Many steps of the collapsing
may lead to unnecessary energy increase and superfluous cover of the
graph.

In fact, we can keep all the incompatible cycles of alternative
mutual-determinations without breaking choices for each step in the
leaf-removal core and deal with them for the final reduced solution
graph. All the backbones have no influence on the solution space,
but breaking the incompatible cycles of alternative
mutual-determinations on the reduced solution graph is an urgent
task for achieving the proper solution subspaces. Many these
incompatible cycles are coupled together and should be broken by
making some nodes on them negatively frozen. The fewer the number of
negatively frozen nodes are chosen, the better covers we can obtain.
Therefore, this problem can be reduced to the MAX-CUT \cite{cut}
problem for the unfrozen nodes of the reduced solution graph. By the
results of MAX-CUT, breaking the edges of unfrozen nodes which do
not belong to the max-cut will lead to totaly compatible cycles.
However, the MAX-CUT problem is also a NP-complete problem which is
hard to solve.

\section{Conclusion and discussion}

A new solution space structure, mutual-determination between
unfrozen nodes is defined and discovered in some detailed case of
the Vertex-cover. Based on the mutual-determinations and backbones,
we construct the reduced solution graph to reveal the solution space
of Vertex-cover. And, inspired by the leaf removal process and
introducing node ranks, a dynamical process for the evolution of the
node states is studied to achieve the current states of nodes in the
reduced solution graph. Combing the mutual-determinations, backbones
and the node ranks, an algorithm named Mutual-determination and
Backbone Evolution Algorithm is proposed to obtained the accurate
reduced solution graph. To ensure the accuracy of the algorithm, the
releasing operations, checking and rechecking techniques and odd
cycles breaking operation are defined by considering the influence
propagation. Then, the numerical results and some examples are given
to verify the validity of the algorithm. Besides, we have proved
that this algorithm is an $O(N^2)$ algorithm and performs strict
when there is no leaf-removal core for the graph. the influence of
incompatible cycles of unfrozen nodes to the algorithm is given,
which can be reduced to MAX-CUT problem.

The Mutual-determination and Backbone Evolution Algorithm can be
applied to a wide range of graphs. Though the difficulties are
brought to by the incompatible cycles in case D on the reduced
solution graph, choosing proper strategies to break the cycles will
be helpful to obtain a solution subspace, which will be beneficial
to solve the Vertex-cover problem in different topologies. Besides,
in order to break the incompatible cycles of unfrozen nodes in case
D on the reduced solution graph, we should design better heuristic
strategies to check the key unfrozen nodes on it, such as taking
advantage of the centrality or clustering coefficient
\cite{centrality,centrality2}. However, as the intrinsic character
of Vertex-cover is NP-complete, the Mutual-determination and
Backbone Evolution Algorithm will still be an approximated one, and
what we aim at is to improve the accuracy of solving different
graphs of Vertex-cover.

 The reduced solution graph of
Mutual-determination and Backbone Evolution Algorithm can correspond
to the whole solution space of Vertex-cover in some cases, which is
a great help to count the number of solutions. Similar as the \#CSP
\cite{count,count2}, the \#Vertex-cover can be analyzed based on the
reduced solution graph. However, calculating the exact entropy of
the solution space needs a much detailed analysis of the
constructions of the reduced solution graph and there should be many
techniques to be introduce on counting the solutions on the reduced
solution graph. Besides, the reduced solution graph can help
explicitly determine the role of every node and calculate the
partition functions and marginal probabilities of the
nodes/variables. Some of the related results will be proceeded in
our future work.

The principal of our algorithm is related to the replica symmetry
theory but not restrict to. Most of recent algorithms solving
combinatorial optimization problems always concentrate to find one
solution, such as the searching algorithms, heuristic algorithms and
even the Belief/Survey Propagation algorithm \cite{algorithms,sp}.
They assign values to the nodes/variables according to some
strategies and do backtracking to reach the optimal solution, or
determine the probability of the variables taking some values. The
Mutual-determination and Backbone Evolution Algorithm collects as
more solutions as possible for the initial subgraphs, and aims to
find solutions by contracting the solution space. At least,
algorithms of detecting the solution space provides a strategy of
reducing the complexity of finding solutions, and combing our
algorithm with other searching and heuristic algorithm may be an
interesting research direction for accelerating the solving process.

\section*{Acknowledgment}

This work is supported by the Fundamental Research Funds for
the Central Universities.\\

\end{document}